\documentclass[showpacs,preprintnumbers,aps,prb,twocolumn,floatfix]{revtex4}

\usepackage[dvips,xdvi]{graphicx}
\usepackage{epsfig}
\usepackage{amsmath}

\begin {document}
\title{Polaritons in cuprous oxide perturbed by LA-phonons}
\author{Oleksiy Roslyak and Joseph L. Birman}
\affiliation{Physics Department, The City College, CUNY \\
Convent Ave. at 138 St, New York, N.Y. 10031, 
USA}
\date{\today}

\begin{abstract}
We present a comparative analysis of a 'conventional' phonoriton (coherent superposition of exciton-photon-phonon ) and a polariton (coherent exciton-photon superposition) 'weakly' coupled to the LA-phonons bath. Depending on duration of the pumping laser field the phonon-induced decoherence results in two distinct types of excitation. Long ($ms$) laser pumping pulses form an 'equilibrium' polariton. The generic feature here is a pronounced photo-thermal bi-stability. i.e. formation of four distinct branches. Transitions between branches can be achieved by excitation energy fluctuations as small as $200 \; neV$ which may impede BEC of the para-excitons. Short ($\mu s$) laser pulses create a 'quasi-equilibrium' polariton. In the latter case, for some critical intensity of the laser field we demonstrate possibility of strong luminicence from a highly unstable localized state on the lower polariton branch. 
\end{abstract}
\pacs{71.35.-y, 71.35.Lk, 71.36.+c}
\maketitle
\section{INTRODUCTION }

Cuprous oxide is considered as a probable candidate for BEC of excitons \cite{SNOKE:1992, FROHLICH:2005, IVANOV:1998,MOSKALENKO:2000}. The extremely small mass of the excitons allows them to condense at temperatures much higher than those of such atomic systems as helium or alkali metals \cite{ANTHONY:1996}. The lowest excited states of  cuprous oxide are classified following the cubic symmetry of the lattice ($ O_h $). This gives the non degenerate ortho-exciton ($ \Gamma_5^+ $,[OE] , binding energy $E_b = 150 \ meV$) and triple degenerate para-exciton ($ \Gamma_2^+ $,[PE] , binding energy $E_b = 162 \ meV$). Due to even parity OE are quadrupole active, and PE remain optically inactive in one photon absorption experiments. However the PE can be effectively populated from the OE by emitting optical phonons or in two photon experiments. For the samples of high purity (more then 99.99 \%) the line-width  of OE is very narrow ($ \Gamma_{1S}\approx 0.8 \ \mu eV $  for $ T=2K $) and detrmined mostly by OE to PE conversion rate. This narrowness is responsible for the extremely long lifetime and profound ($ns$) BEC time coherence for the PE \cite{SNOKE:1990}. 
\par
There are two main mechanisms impeding expected BEC in bulk cuprous oxide samples and both have non-statistical origin. First is Auger heating \cite{KAVOULAKIS:1996}. It was demonstrated that although OE run along an adiabat slightly above the condensation curve the PE get condensed. Second impeding mechanism is the relatively strong coupling to photons, so that such polaritons tend to escape the crystal before condensation occurs \cite{IVANOV:1998}. For such polaritons to experience BEC, there is a criterion of smallness imposed on the exciton-photon coupling. Only PE meet the criteria \cite {FROHLICH:2005}, but as far as we know there is no conclusive evidence of BEC in the bulk cuprous oxide. 
\par
In this paper we examine another possible mechanism obstructing BEC. Namely, a weak coupling of the polariton branches with long wavelength acoustical phonons and the effective formation of a new type of polariton. For low pumping intensity and in close proximity to the exciton-photon resonance one can consider the 'weak' coupling to the LA-phonons \footnote{The temperature is low enough to neglect contributions from the optical phonons} and treat the phonons semi-classically by means of some effective temperature dependence of the exciton energy. Namely, in the case of weak (but not negligible) electron-phonon coupling the semiconductor gap energy (and thus the  energy of the exciton) decreases. There is experimental evidence of the exciton energy red-shift for various semiconductors \cite{MYSYROWICZ:1979} as well as for $ Cu_2O $ at helium temperatures \cite {DASBACH:2004}.
\par
In section II we compare this weakly perturbed polariton versus the concept of conventional phonoriton as an excitation formed by resonance interaction between the exciton, photon and phonon \cite{BIRMAN:1990}. The resulting polariton branches strongly depend upon what kind of heating  and cooling mechanism dominates. 
\par
In  section III we investigate the pumping of a cuprous oxide thin film by short and ultrashort laser pulses ($\mu s$). In this regime the heating is a result of 'phonon assisted' Auger process \cite{KAVOULAKIS:1996} and the cooling is provided by LA-phonons. The polaritons come to quasi-equilibrium and due to the temperature dependence of the exciton energy the dispersion is modified. Consequently we show localized states on the lower branch of this new polariton. 
\par  
In  section IV we regard long laser pulses ($ms$). As a consequence, the temperature is defined by the laser heating and temperature exchange with the surrounding helium. We demonstrate that this results in temperature bi-stability and splitting of the dispersion curves into four distinct branches. 
\par
Numerical estimates of the decoherence effects are performed in the concluding section V using realistic material parameters. According to our calculations this splitting and phonon induced decoherence may effectively obstruct BEC of the PE in cuprous oxide.
\section {polariton weakly perturbed by LA phonons vs. conventional phonoriton}
Let us consider a thin film of cuprous oxide ($d_s = 30 \; \mu m$) placed in a helium bath at temperature $T_{bath}=1.7 \; K$ and pumped by laser pulses of varying power $P_{ex}$. The pumping laser field is polarized along ($1\bar{1}0$) and has the wave vector $\bf{k}$ parallel to the ($111$) direction. The laser is tuned into resonance  with dipole forbidden , quadrupole allowed $1S$ exciton ($\hbar \omega_{1S,0} = 2.05 \; eV $ for OE and $\hbar \omega_{1S,0} = 2.05 -0.097 \; eV $  for PE). The exciton mass and radius are $M=3m_0$, $ a_B \approx 5.1 \; \AA$. From the selection rules it follows that only the OE  $(1/\sqrt2)\left(\Gamma^+_{5x}-\Gamma^+_{5y} \right)$ are optically allowed with the oscillator strength of the transition $f_{[111]}=3.7\times10^{-9}$. 
The PE acquire some oscillator strength $ f_p = 4 f_o 10^{-3} $ in a magnetic field or under an external stress through spin-orbit interaction.
\par
The conventional phonoriton \cite {HANKE:1992,IVANOV:1982,BIRMAN:1990,KELDYSH:1986}  is a quasilinear mode depending parametrically on the intensity of the pump light ($P_{ex}$) and characterizes a coherent superposition of exciton, photon , and phonon when up-conversion process is allowed. The laser pulse induces a pump polariton with the exciton wave vector $\bf{p}$ and energy $\hbar \omega_{\bf p}$. The excitonic component of the polariton couples with other excitonic modes $ \bf{k} $, due to the exciton-LA-phonon interaction. To form the standard 'phonoriton', the laser intensity should be big enough to have the exciton-LA-phonon coupling:         
\begin{equation}
\label{EQ:2_2_1}
Q_{\bf k-p } = D_{ex} \sqrt {\frac{{n_o \left| {{\bf{p}} - {\bf{k}}} \right|}}{{2\hbar \rho \upsilon _{ph} }}}  \sim \sqrt {\frac{{P_{ex} }}{{\upsilon _g }}}  
\end{equation}
comparable with the exciton-photon interaction $ \Omega _{rabi} = \sqrt{f_o}E_{1S} \approx 124 \; \mu eV $. Here we introduced $ D_{ex} $ as the exciton deformation potential; $ \rho $ is the mass density; $ \upsilon_{g} \ , n_0 $ are the group velocity and density respectively of the polariton.
In terms of creation (annihilation) operators, the exciton ($b^\dag_{\bf k}$) - photon ($c^\dag_{\bf k-p}$) - phonon ($\alpha^\dag_{\bf k}$) Hamiltonian of the cuprous oxide crystal has form:
\begin{equation*}
\begin{split}
H/\hbar = \left({\omega_{1S, {\bf k}}-\omega_{\bf p}}\right)b^\dag_{\bf k}b_{\bf k}+\left({\omega^{photon}_{\bf k}-\omega_{\bf p}}\right)\alpha^\dag_{\bf k}\alpha_{\bf k}+\\
\Omega_{\bf p-k}^{phonon} c^\dag_{\bf k-p}c_{\bf k-p}+\frac{i}{2} \Omega_{rabi,{\bf k}}\left({\alpha^\dag_{\bf k}b_{\bf k}-b^\dag_{\bf k}\alpha_{\bf k}}\right)+\\
Q_{\bf k-p}\left({b^\dag_{\bf k}c_{\bf k-p}-c^\dag_{\bf k-p}b_{\bf k}}\right)
\end{split}
\end{equation*}
Here $\omega^{photon}_{\bf k}=ck/\sqrt{\epsilon_\infty}$ and $\Omega_{\bf p-k}^{phonon}=v_s \left|{\bf k-p}\right|\approx 155 \; \mu eV $ are the photon and phonon dispersion at the exciton-photon resonance $k_0=2.62\times10^5 \; cm^{-1}$; $\epsilon_\infty = 6.5$ is the background dielectric constant; $v_s$ is the LA-sound velocity. The eigenvalues of the Hamiltonian above are given by an implicit phonoriton dispersion:
\begin{equation}
\label{EQ:2_2_2A}
\begin{split}
\left( {\omega _{1S,{\bf k}} - \omega } \right)\left( {\omega _{\bf{k}}^{photon}  - \omega } \right)\left( {\omega _{\bf{p}}  + \Omega _{{\bf{k}} - {\bf{p}}}^{phonon}  - \omega } \right)-\\
Q_{{\bf{k}} - {\bf{p}}}^2 \left( {\omega _{\bf{k}}^{photon}  - \omega } \right) - \frac{{\Omega _{rabi,{\bf k}}^2 }}{4}\left( {\omega _{\bf{p}}  + \Omega _{{\bf{k}} - {\bf{p}}}^{phonon}  - \omega } \right) = 0
\end{split}
\end{equation}
In contrast to the standard picture of phonon-mediated polariton scattering, the dispersion is strongly modified from the polariton dispersion and consists of three dispersion branches. Around the exciton-photon resonance, such coherent polariton-phonon interaction results in an effective 'blue' shift of the 'polariton-like' branch of the phonoriton. 
\par
The key point in the phonoriton picture is the coherence between all the particle involved. However for small density of the polaritons the polariton-phonon interaction (\ref{EQ:2_2_1}) is 'weak'. Therefore, we define the 'weak' polariton -phonon coupling as a limiting case of conventional phonoriton when the down-conversion \footnote{Polariton $\to$ phonon bath} processes are dominant and up-conversion \footnote{Phonon bath $\to$ polariton} processes are negligible. Therefore, in close proximity to exciton-photon resonance this 'weak' interaction with the LA-phonons bath manifests itself by small decoherence between exciton and photon in the polariton. 
\par
We propose to treat the LA-phonons semi-classically and use deviation of the polariton gas temperature from the temperature of the helium bath $T-T_{bath}$ as that small decoherence parameter. To estimate the influence of the temperature deviation on the polariton dispersion, we adopt the following picture of the exciton-phonon interaction. 
\par
The electron and hole of the exciton part of the polariton move in the cloud of  LA-phonons (we neglect TA phonons, as their contribution to the interaction is much weaker at the given temperature range \cite{KAVOULAKIS:1996} ). The result is a bigger mass of the exciton and red shift of the \emph{energy gap} between the conduction and valence bands without noticeable effect on the binding energy of the $1S$ exciton \cite{SNOKE:1990}. There is experimental evidences for such exciton-phonon interaction. Exciton energy 'red' shift has been observed for several types of semiconductor \cite{MYSYROWICZ:1979} as well as specifically for $ Cu_2O $ at helium temperatures \cite {DASBACH:2004}. 
\par
According to general formulation of the theory \cite{PASSLER:1998} this temperature dependence is accurately given by  integrals of the form:
\begin{equation}
\label{EQ:2_2_4}
\begin{split}
\omega_{1S,0} \left({T}\right) - \omega _{1S,0} \left({T_{bath}}\right) = \\
-\frac{1}{2}\int {d\omega } f\left( \omega  \right)\left( {\coth \left( {\frac{{\hbar \omega }}{{2k_B \left({T-T_{bath}}\right)}}} \right) - 1} \right)
\end{split}
\end{equation}
Here $ {\coth \left( {\frac{{\hbar \omega }}{{2k_B \left({T-T_{bath}}\right)}}} \right) - 1} $ represents the average phonon occupancy number and $ f(\omega) $ is the relevant electron-phonon spectral function.
\par
\begin{figure}[htbp]
	\includegraphics{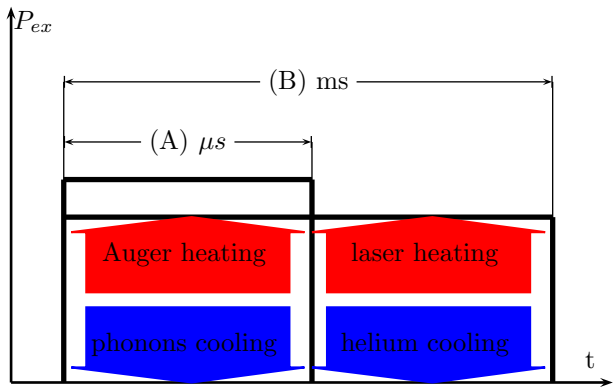}
	\caption{(Color on-line) Schematic of different heating (cooling) mechanism in the thin film of cuprous oxide. Different mechanisms create different thermal regime of the sample and result in different polariton dispersions. The case (A) corresponds to short excitation pulses, $T$ is the temperature of the exciton gas, $T_{bath}$ is the temperature of the LA phonons; and the case (B) corresponds to long pulses, $T$ is the temperature of the LA phonons, $T_{bath}$ is the temperature of the surrounding helium.}
	\label{FIG:2_1}
\end{figure}

Expanding (\ref{EQ:2_2_4}) into series around the bath temperature, in linear approximation the dispersion of this perturbed polariton is given by the following system of equations:
\begin{eqnarray}
\label{EQ:2_2_2}
\left( {\omega _{1S,{\bf k}}\left( T \right)  - \omega } \right)\left( {\omega _{\bf{k}}^{photon}  - \omega } \right) - \frac{{\Omega _{rabi,{\bf k}}^2 }}{4}= 0\\
\nonumber
\hbar \omega _{1S,{\bf k}}^{}\left( T \right)= \hbar \omega _{1S,0} \left({T_{bath}}\right)+\frac{\varepsilon _\infty \hbar^2 k^2}{2 M} + \\
\nonumber
+\kappa \left( {T - T_{bath} } \right)
\end{eqnarray}
In the last expression the parameter $\kappa = 0.30 \ \mu eV \ K^{-1}$ is experimentally determined \cite{DASBACH:2004}. The resulting polariton branches are given by the equations  above along with a temperature equation which is strongly dependent upon what kind of heating and cooling mechanism  prevails (See Fig.\ref{FIG:2_1}). 
Therefore in the next sections we are going to consider two distinct cases:
\begin{itemize}
	\item [(A)] Short ($\mu s$) laser pulses, where heating is a result of 'phonon-assisted' Auger process \cite{KAVOULAKIS:1996} and the cooling is caused by LA-phonons. In  pico-second dynamics the system comes to quasi-equilibrium, and due to the temperature dependence of the excitonic part of the polariton there is a distinct change in the dispersion. 
	\item [(B)] Long ($ms$) laser pulses are applied, which leads to the temperature being defined by photo-thermal laser heating and temperature exchange with the surrounding helium. This will result in a photo-thermal bi-stability effect.
\end{itemize}
Note that the 'polariton-like' branches of the phonoriton given by equation (\ref{EQ:2_2_2A}) can be formally obtained from the system  (\ref{EQ:2_2_2}) by modeling the up-conversion processes with the additional condition $T_{bath}>T$. But the physical meaning of the system (\ref{EQ:2_2_2}) is rather different. In our treatment the phonons are not mediators for the exciton-exciton interaction but the source of decoherence for the pump polariton.
\section{Short laser pulses}
First we consider short laser excitations (See Fig.\ref{FIG:2_1}A). Therefore the system of interacting $1S$ quadrupole excitons, photons and low energy LA phonons can be treated adiabatically.
\par
In the equation governing the dynamics of the temperature we neglect all ultra-fast processes (pico and nano second dynamical processes), such as optical phonon cooling ( characteristic time $ \tau  = 0.2 \times T^{ - 1/2} \ ns $ ); radiative decay ( $ \approx 300 \ ns $); ortho-to-para down conversion ( $ \approx 0.014 \times T^{3/2} \ ns^{ - 1} $ )\cite{KAVOULAKIS:1996}.
\par
The main sources of heating in this adiabatic system are the non radiative direct process and phonon assisted Auger process with the characteristic Auger decay time  $ \approx 0.1 \times T^{-3/2}\  \mu s $. Because the conduction and valence band share the same parity, the rate of direct Auger process is negligibly small. Hence, one has to consider 'phonon-assisted' Auger process which is defined as exciton-exciton recombination into hot electron hole carriers accompanied by emission of a cascade of phonons (See Ref.\cite{KAVOULAKIS:1996} for more complete discussion). 
\par
The rate of change of the entropy of the excitons is a balance between the entropy loss due to phonon cooling and heating following Auger annihilation of the excitons. The rate of phonon cooling, which is dominated by emission of acoustic phonons, varies as $-a_1 T^{3/2} \left( T-T_{bath} \right)$. On the other hand, heating of the excitons from the Auger process is proportional to the density of the polaritons $a_{2} n_0$ \footnote{For the OE it comes close to the condensation line $n_0 \propto T^{3/2}$ at high density of the exciton gas but never crosses. For the PE there is BEC critical temperature given by the cross point.}. The ratio of the coefficients $a_1/a_2$ varies from $4.75\times 10^{22} $ to $5.71\times 10^{22} $ and rapidly grows with increase in duration of the pulses. This emphasizes the fact that for long pulses one can neglect the effect of the Auger heating comparing to the phonon cooling and has to consider direct heating by the laser field. Quantative comparison of the thermal red shift to other high density effects reported so far for the cuprous oxide crystals can be found in the Appendix. 
\par
Consequently, with increasing temperature, the exciton density grows along the 'quasi-equilibrium' adiabat:
\begin{eqnarray}
\label{EQ:2_3_1}
\ -a_{1} T^{3/2} \left( T-T_{bath} \right)+ a_{2} n = 0 \\
\nonumber
n\left({\omega,T}\right) = \frac{4P_{ext}}{\pi d^2 \hbar \omega_{1S} v_g } \\
\nonumber
v_g^{ - 1}  = \frac{{\sqrt {\varepsilon _\infty  } }}{c}\frac{{\partial x}}{{\partial \hbar \omega}} \approx \frac{{\sqrt {\varepsilon _\infty  } }}{c}\frac{x-\hbar \omega}{\hbar \omega_{1S,x}\left({T}\right) - \hbar \omega}
\end{eqnarray}
Here $ x = \hbar ck / \sqrt{\epsilon_{\infty}} $ is the energy of the pumping laser field.
\par
\begin{figure}[htbp]
	\centering
	\includegraphics[width=8.6 cm]{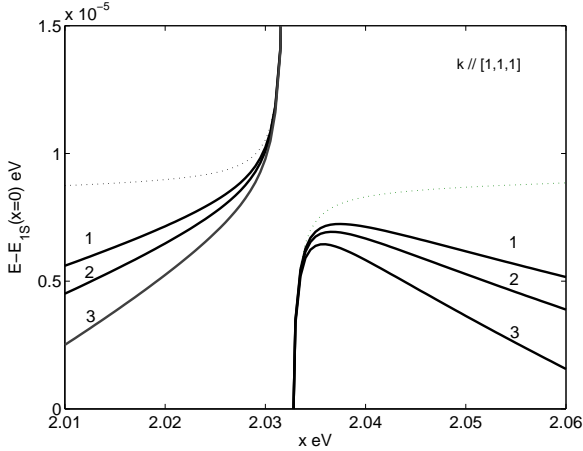}
\caption{Polariton dispersion for short pump laser pulse and quasi-equilibrium of the exciton gas with the phonon bath. The 'phonon-assisted' Auger process heats the exciton gas and weak interaction with LA-phonons cools it down. Solid lines represent two polariton branches for varying laser intensity: $ \left({1}\right) \; P_{ex} = 5 \ mW $; $ \left({2}\right) \; P_{ex} = 10 \ mW $; $ \left({3}\right) \; P_{ex} = 25 \ mW $. The dash line corresponds to the unperturbed polariton: $ P_{ex} < 5 \ mW $.}
	\label{FIG:2_2}
\end{figure}

The system of equations (\ref{EQ:2_3_1}, \ref{EQ:2_2_2}) has only one solution for the temperature (no bi-stability). Therefore, it can be numerically resolved into the upper and lower branches of the perturbed polariton. To specify the branches for the phonoriton we put the following conditions for the upper and lower branch correspondingly: 
\begin{equation}
\label{EQ:2_3_2}
\mathop {\lim }\limits_{x \to x_0^ + ,x \to x_0^ -  } \frac{\hbar}{x}\frac{{\partial \omega }}{{\partial x}} = 1
\end{equation}
In close proximity to the exciton-photon resonance ($x_0 = \hbar \omega_{1S}$) and for small light intensities, the solution of the nonlinear system (\ref{EQ:2_3_1},\ref{EQ:2_2_2}) can be obtained by perturbation theory. Therefore, in first order, we set an appropriate (\ref{EQ:2_3_2}) unperturbed group velocity into the second equation of the system (\ref{EQ:2_3_1}) and solve the corresponding algebraic system of equations (\ref{EQ:2_2_2}) for new upper and lower branches of the polariton. The resulting polariton dispersion ($E = \hbar \omega$) as a deviation from the exciton energy ($E_{1S,x} = \hbar \omega_{1S,{\bf k}}$) is presented in Fig.\ref{FIG:2_2}.
\par
For high laser intensities ($P_{ex} > 25 \ mW$) the first approximation is not enough, and one must carry on with more iterations \footnote{As an alternative to iterations one can approach the desirable intensity by changing it gradually.} which reduces the red shift effect, as the group velocity grows. 
\par
The physics behind different levels of the perturbations can be described as the following. In zero approximation, a short laser pulse of energy close to the resonance $ x_0 $ enters the system and creates a polariton wave. The temperature of the system is equal to the $ T_{bath} = 1.7 \ K $ of the surrounding. 
\par
\begin{figure}
	\centering
	\includegraphics[width=8.6 cm]{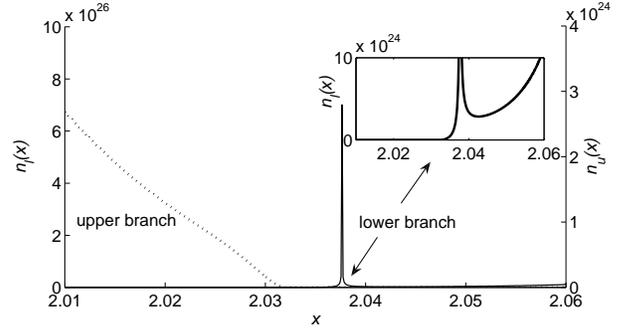}
	\caption{Density of polariton upper ($n_u$) and lower ($n_l$) branches in first approximation which must be used for the second approximation in the perturbation. The density peak represents the polaritonic build up on the lower branch as the result of high localization ($v_g \to 0$).}	
	\label{FIG:2_4}
\end{figure}

In first approximation, the polariton is relaxed due to polariton-exciton interaction. On the exciton-like part of the polariton (away from $ x_0 $) this leads to  heating due to Auger process and as a consequence the 'red' shift of the exciton energy. The perturbed polaritons accumulate on the lower branch according to the new group velocity. The polariton density build-up ($ P_{ex} = 5 \ mW $) is illustrated in Fig.\ref{FIG:2_4}.
\par
In the second approximation, this energy change increases the group velocity of the polariton waves and the exciton density is reduced and so is the temperature of the polariton gas. Therefore the polariton branches swing slightly back to the unperturbed polariton dispersion. Presumably, the quasi equilibrium between the Auger heating and LA-phonon cooling is established after some more steps. 
\section{Long laser pulses}
Now let us consider the case of long ($ms$) pumping signals. Contrary to the adiabatic case of short pulses, the heating is due to the laser radiation and cooling is due to surrounding helium (See Fig.\ref{FIG:2_1}B) for the open system. As we are about to demonstrate, it results in bi-stability of the temperature and splitting of the dispersion curves into four different branches. 
\par
The bi-stability effect is due to the nonlinear origin of the temperature equation \cite{DASBACH:2004}. The heating process is proportional to the absorption  inside the semiconductor film and therefore strongly dependent on the polariton dispersion and proximity to the exciton-photon resonance. Neglecting the reflection, one can define the absorption through the temperature dependent transmission as: $1-{\rm Tr \left({T}\right)}$. Therefore, the equilibrium temperature is given by the stationary solution of the following time $\tau$ dependent system of equations: 
\begin{eqnarray}
\label{EQ:2_4_1}
\frac{{dT}}{{d\tau }} = HP_{ex} \left[ {1 - {\rm Tr}\left({\hbar \omega, T }\right)} \right] - C\left[ {T - T_{bath} } \right] \\
\nonumber
{\rm Tr} \left({\hbar \omega, T }\right) = \exp \left[ { - \frac{{\Gamma _{1S}^2 }}{{4\left( {\hbar \omega_{1S} \left( x,T \right) - \hbar \omega} \right)^2  + \Gamma _{1S}^2 }}\frac{{d_s }}{{l_a }}} \right]
\end{eqnarray}
Here the phenomenologically introduced constants $ H=5.0\times 10^6 \ K s^{-1} W^{-1} $ and $ C=2.8\times 10^{3}\  s^{-1} $ giving the heating and cooling rate, respectively; $\Gamma_{1S} = 0.8  \; \mu eV$ is the phenomenological line-width; the absorption length is given by:
\[
l_a  = \frac{{\Gamma _{1S} \hbar c\sqrt {\varepsilon _\infty}}}{f_o \hbar \omega_{1S, x_0}^2}
\]
The factor $1/4$ counts single-degenerate OE and triple-degenerate PE.
\par
\par
For a qualitative description of the bi-stability we fix the wave vector at energy $ x = \hbar \omega = 2.01 \; eV $ slightly off the resonant value $x_0$ and consider only the upper branch of the perturbed polariton. Then the heating part $H P_{ex} \left[ {1 - {\rm Tr}\left( \hbar \omega, T  \right)} \right]$ of the temperature equation (\ref{EQ:2_4_1}) has a Gaussian shape, with the maximum dependent on detuning from the polariton energy. Depending on the value of the detuning it can cross the cooling line $C\left[ {T - T_{bath} } \right]$ on the phase diagram ($\partial T / \partial \tau$ vs. $T$) in one, two or three points on the the temperature scale. The last case corresponds to a strong nonlinearity and a hysteresis loop for the temperature in momentum-energy space.
\par
\begin{figure}[htbp]
	\includegraphics[width=8.6 cm]{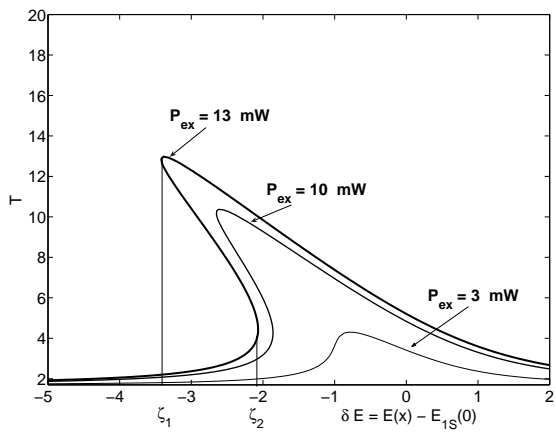}
	\caption{The first equation of the system (\ref{EQ:2_4_1}) is visualized as an inexplicit function of the local sample temperature $T$ and detuning $\delta E$ from the resonance energy for different excitation intensities $ P_{ex} $. The red shift of the maximum is accompanied by the bi-stability effect.}	
	\label{FIG:2_5}
\end{figure}

If one moves from higher to lower energy (detuning $ \hbar \omega (x)-E^{up}>0 $), temperature increases as the transmission decreases (See Fig.\ref{FIG:2_5}). The shift of the excitonic part of the polariton and low transmission extend to large negative detuning and allows  efficient heating. If one starts from  negative values for the detuning, the sample is cold even beyond the ' cut-off ($ \zeta $)' energy. Transmission remains close to unity and photo-thermal heating is inefficient. This means that the equilibrium state for the cold (up-going) scan and already heated (down-going) scan are different. Let us consider the down-going scan. When one closes up to the polariton upper branch energy the transmission decreases - which increases the sample temperature. At some point ('cut-off' energy) the heating becomes dominant over the cooling and we face a rapid temperature drop. The range of bi-stability mainly depends on laser pulses intensity $ P_{ex} $. 
\par
\begin{figure}[htbp]
	\includegraphics[width=8.6 cm]{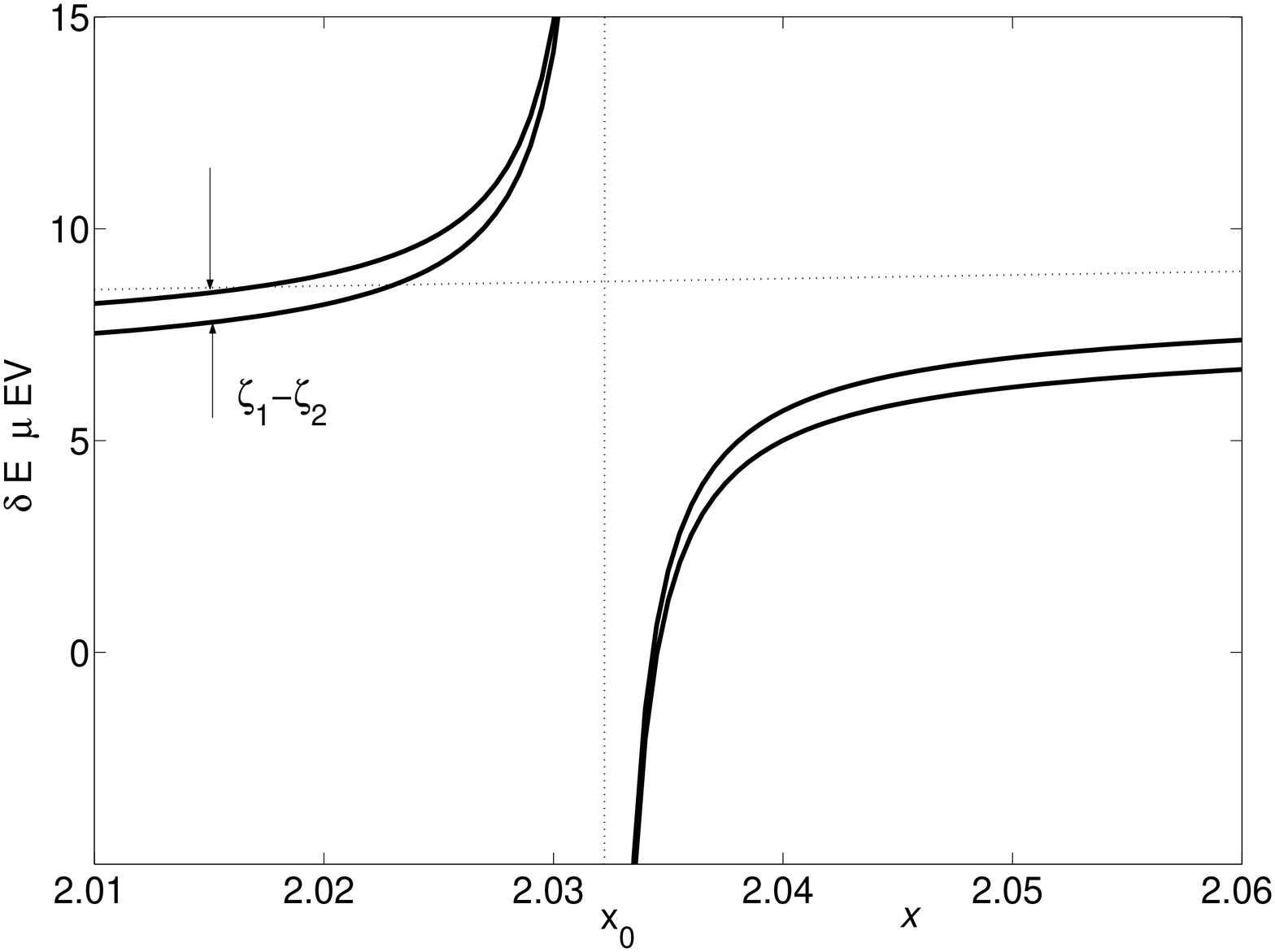}
		\reflectbox{\includegraphics[width=7 cm]{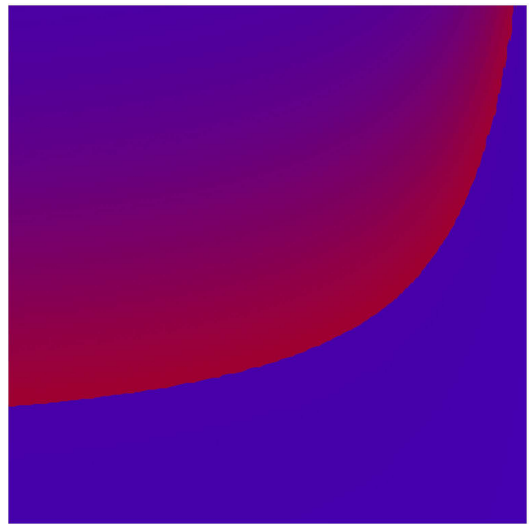}}
	\caption{(Color on-line) The resulting branches (solid) of the quadrupole-LA phonoriton dispersion and corresponding temperature profile (upper branch only). The dashed lines correspond to noninteracting photon (vertical) and 1S quadrupole exciton (horizontal). The laser intensity is taken to be $ P_{ex} = 5 mW $. The bi-stability range for the exciton-like part of the phonoriton is approximately $ 2 \mu eV $. Actual population of the branches depends on the initial condition and up or down-going scan. Also in case of big light intensity $ P_{ex} >57 mW $ interconversion processes can occur due to optical $ \Gamma^-_{12} $ phonons }	
	\label{FIG:2_6}
\end{figure}

For a quantative numerical calculation of the resulting dispersion curves, one has to resolve the nonlinear system of equations (\ref{EQ:2_2_2}, \ref{EQ:2_4_1}) at equilibrium using standard perturbation theory. As a result of nonlinear origin of the temperature equation, there are four different polariton branches corresponding to the up- and down-scan (See Fig.\ref{FIG:2_6}). The 'bi-stability' effect manifests itself as abrupt drop in the temperature from red color for a large temperature to blue color for the cold helium.

\section{Results and discussion}  
The most important result which emerges from the dispersion curves in case of pumping by short pulses  is the existence of a \emph{localized state} on the lower branch of the polariton (Fig.\ref{FIG:2_4}). So on a $ \mu s $ time interval one can observe replicas of these states in the luminescence spectra due to intense polariton build up in this state. 
\par
The evidence of such polariton density build-up modes can be demonstrated in the collapse and revival of the quantum beats of the scattered probe signal \cite{FROHLICH:1991}. The luminescence from this highly unstable state can be considered as a signature of the polariton perturbation by the LA-phonons and can be utilized to produce delayed laser pulses. The states on the upper branch have bigger group velocity compared to standard polariton modes, and this can effectively prevent BEC as the polaritons escape the crystal more effectively.
\par 
In the case of long pump pulses an even more effective mechanism which prevents BEC is provided by the bi-stability effect. Indeed it was demonstrated by Ell \cite{IVANOV:1998} that the BEC for excitation of states with sufficiently low oscillator strength is possible only when the kinetic energy of the excitonic part of the pump polariton is bigger than the splitting between the polariton branches at  resonance. The latter prevents the polariton from 'condensing' outside of the crystal. This condition  certainly fails for the OE polaritons, since the kinetic energy is about $ 9 \; \mu eV $ and the Rabi splitting is $127 \; \mu eV $. Moreover, the adiabat (\ref{EQ:2_2_2}) never crosses the condensation line $ T_c = n_0^{2/3} \times 10^{-11} $.
\par 
However, PE are good candidate for possible BEC and live long enough ($ > 10\; ms $) to reach the equilibrium because the Rabi splitting is $ \hbar\Omega_p = \hbar\Omega_o \sqrt{f_p/f_o} = 1.1\; \mu eV $ for $ 1 \; T $ magnetic field, which is eight times less then the kinetic energy. For rather large intensity of the pumping laser,according to the results of this paper, the polaritons are perturbed by the 'weak' interaction with LA-phonons. Hence, due to the big life time of the para-exciton  one has splitting of the upper and lower polariton branches into four new ones due to bi-stability effect. The extremely small oscillator strength for this para-exciton transition $ f_p = 4 f_o \times 10^{-3} $ supports the condensation condition stated above for the films starting from $ d_s > 0.7 \; mm $ thickness \cite{KAPINSKA:2005}. 
\par
The upper and lower PE polariton branches are split into four new ones due to bi-stability effect in close analogy to the OE polariton. Let us estimate the effective splitting between all pairs of the polariton branches. According to the second equation of the system (\ref{EQ:2_2_2}), to reach the critical condensation density of $ n_c \approx \left(9 T_{bath} \times 10^{11} \right)^{3/2} = 1.89 \times 10^{18} \; cm^{-3} $ at the point of resonance with light, one needs intensity of the laser field $ P_{ex} = 185 \; mW $. In this case the calculated region of  bi-stability is about $ \zeta_1 - \zeta_2 = 25 \; \mu eV $ which is approximately two times bigger than the kinetic energy of the PE. As a result BEC of the PE polaritons may be effectively \emph{suppressed} by the bi-stability effect. 
\par
In conclusion we would like to briefly outline possible deviations from the linear model of the temperature dependent exciton energy in cuprous oxide crystals which we used. If one has to consider the influence of the bi-stability effect on the possible BEC of the polariton or for high intensity of the pumping laser $ P_{ex} > 35 mW $ when the temperature growth is bigger than $87 \; K$ then the linear approximation for the temperature dependence of the gap energy given by (\ref{EQ:2_2_2}) is may be not applicable.
\par
Therefore, one has to apply more elaborate expression for the exciton energy $E_{1S} \left( T,x \right) \to E_{1S} \left( T,x,\rho \right)$ including a contribution from the optical phonons \cite{PASSLER:1998,MYSYROWICZ:1979}.  The parameter $ \rho $ controls the relative contribution of the long-wavelength acoustic phonons $ \rho\to 1 $ , on one hand, and those of optical and short-wavelength acoustical phonons, on the other hand $\rho \to 0$. The dielectric background constant $\epsilon_\infty $ becomes a linear function of the wave vector.
\par
Due to the complexity of the full expression it is useful to consider  some approximations. When the acoustic phonons dominate, this expression transforms into the familiar Bose-Einstein model proposed by Vina \cite{VINA:1992}:
\begin{equation*}
\hbar \omega_{1S} \left( T \right) = \hbar \omega_{1s} \left({0}\right) - a\left[ {1 + \frac{2}{{exp \left({\Theta / T} \right) - 1}}} \right]
\end{equation*}
Here there are two fitting parameters  $ a $ and $ \Theta $. Both emission and absorption of phonons are now considered. For the LA phonons the main contribution to the exciton energy can be fitted with the empirical relation proposed by Varshi \cite{VARSHI:1967}: 
\begin{equation*}
\hbar \omega_{1S} \left( T \right) = \hbar \omega_{1s} \left({0}\right) - \frac{{\alpha T^2 }}{{T + \Theta }}
\end{equation*} 
\par
Experiment \cite{DASBACH:2004} shows that for the bulk cuprous oxide $\alpha = 4.8\times 10^{-7} \ eV/K $.
 
\begin{acknowledgments}
We would like to acknowledge Ms. Upali Aparajita for useful comments and discussion on the manuscript. This project was supported in part by PCS-CUNY.  
\end{acknowledgments}
\bibliography{bibliography} 
\appendix
\section{}
The comprehensive review article of Fernandez-Rossier \cite{FERNANDEZ:2006} deals with exciton high density effects as the density approaches saturation. Here we apply their theory to bulk cuprous oxide. The $1S$ exciton energy experiences 'blue' shift due to exchange interaction between electrons and holes and red shift due to increased screening:
\begin{eqnarray*}
\Delta E^{blue}  = \frac{{13\pi }}{3} \hbar \omega_{1S,0} n a_B^3\\
\Delta E^{red}  = f_{o} \left( \omega  \right)\hbar \omega_{1S,0} \pi n a_B^3 
\end{eqnarray*}
The screening spectral function is given by the following expression:
\[
f\left( \varpi  \right) = \frac{{\varpi \left( {32 + 63\varpi  + 44\varpi ^2  + 11\varpi ^3 } \right)}}{{\left( {1 + \varpi } \right)^4 }} - \frac{{8\varpi \left( {4 + 3\varpi } \right)}}{{\left( {1 + \varpi } \right)^2 }}
\] 
and $\varpi ^2  = \left( {1 + m_h /m_e } \right)^2/\left({4 m_h / m_e }\right)$. 
These effects partially compensate each other. 
\par
To estimate the net shift let us take for example the following numerical vales: $ P_{ext} = 5 \ mW $ is the laser field intensity; $ d = 1\times 10^{-6} \ m $ is the diameter of the focused laser beam; $ \upsilon_g = 4\times 10^4 \ m/s $ represents the group velocity of the pump polariton, and the factor $4$ indicates the fact that we excite both OE and PE; $m_e  = 0.69 m_0 $ and $m_h  = 0.99 m_0 $ are the effective mass for electron and hole.
\par
Due to small quadrupole exciton radius the net shift: $\Delta E^{red}-\Delta E^{blue} = - 0.226 \; \mu eV$. Hence, it is much less compared to the red shift due to the heating: $ - 2.1 \; \mu eV $. But the situation will change drastically for a 2D quantum well, as in this case the blue shift is  dominated due to exchange energy. Note that these estimations are valid for any duration of the pulses required to build up the necessary polariton density.
\par
The oscillator strength is proportional to the gap energy of cuprous oxide \cite{MOSKALENKO:2002} and so depends on the temperature also . It gets smaller with increase of the pumping power. The relative change of the oscillator strength is $ \approx 1.4 \times 10^{-3} $ for $ T-T_{bath} = 20K $, but because it is not a resonant term we neglect this effect later on. For the same reason we neglect the temperature dependence of the line-width \cite{DASBACH:2004}.
\end {document}